\documentclass[11pt]{revtex4}
\usepackage{amsthm,amsfonts,amssymb,epsfig,graphics,amsmath}
\usepackage[latin1]{inputenc}

\numberwithin{equation}{section}

\newcommand{\RR}{{\mathbb R}}

\newcommand{\ZZ}{{\mathbb Z}}

\def\ii{\mathrm{i}}     % imaginary unit
\def\eps{\varepsilon}

\newcommand{\bspm}{\left(\begin{smallmatrix}}\newcommand{\espm}{\end{smallmatrix}\right)}
\newcommand{\bpm}{\left(\begin{matrix}}\newcommand{\epm}{\end{matrix}\right)}

\newcommand{\PROOF}{\textbf{Proof.} }

\def\epsilon{\varepsilon}

\def\beq{\begin{equation}}
\def\eeq{\end{equation}}

\begin{document}

\title{Vortex families near a spectral edge in the Gross-Pitaevskii equation
with a two-dimensional periodic potential}

\author{Tom\'{a}\v{s} Dohnal$^1$ and Dmitry Pelinovsky$^2$\\
{\small $^{1}$ Fakult\"at f\"ur Mathematik, Karlsruhe Institute of Technology, 76131 Karlsruhe, Germany} \\
{\small $^{2}$ Department of Mathematics, McMaster University, Hamilton, Ontario, Canada, L8S 4K1} }

\date{\today}

\begin{abstract}
We examine numerically vortex families near band edges of the Bloch wave spectrum
in the Gross--Pitaevskii equation with a two-dimensional periodic potential and
in the discrete nonlinear Schr\"odinger equation. We show that
besides vortex families that terminate at a small distance from the band edges
via fold bifurcations there exist vortex families that are continued all way to the band edges.
\end{abstract}

\maketitle

\section{Introduction}

Many physical problems in periodic media with the Kerr (cubic) nonlinearity are
governed by the Gross--Pitaevskii equation with a periodic potential. Examples are
Bose--Einstein condensates in optical lattices \cite{Pitaevskii}
and photonic-crystal fibers \cite{YangSkor}. Interest in properties
of localized states in this model has stimulated a number of mathematical
works devoted to this subject \cite{PelBook,YangBook}.

An interesting problem that arises in this context is the possibility of bifurcations
of stationary localized states from edges of Bloch bands in the wave spectrum
of the Schr\"{o}dinger operator with a periodic potential. First pioneering works
in this direction were completed by C. Stuart and his students \cite{Heinz,Kupper,Stuart}.
In the physics literature the asymptotic approximations of gap solitons bifurcating from
band edges were developed by various authors in one dimension
\cite{PelKivSukh,Hwang} and two dimensions \cite{Shi,ShiWang,Ilan}.

In one dimension it was discovered numerically in \cite[Chapter 6.2]{YangBook} and explained analytically in \cite{Akylas} that while single-pulse gap solitons bifurcate continuously from band edges, double-pulse
gap solitons do not bifurcate from the band edges but experience
fold bifurcations at a small distance from the edges. The situation becomes
even more interesting in the space of two dimensions, where
besides gap solitons, vortex solutions are possible in periodic potentials \cite{DPS09,DU09,Wang}.
However, a contradiction arises between the analytical results of \cite{DPS09,DU09}
suggesting a continuous family of the fundamental vortex solutions bifurcating from the band edges
and the numerical results of \cite{Wang} (see also \cite[Chapter 6.5]{YangBook})
suggesting a fold bifurcation of vortex families at a small distance from the band edges.
This contradiction will be inspected in this paper by using numerical computations.

We will show that there do exist continuous families of the fundamental vortex solutions
bifurcating from band edges, according to the theory in \cite{DPS09,DU09}. Numerical
approximations of these families near band edges suffer, however, from the fact that
the vortex localization is too broad and hence extends beyond the chosen computational domain.
As a result, a spurious fold bifurcation occurs for the fundamental vortex family before
the family reaches the band edge. If the size of the computational domain is enlarged,
the location of the spurious fold bifurcation moves closer to the band edge. At the
same time there are other vortex families, found also in \cite{Wang}, which feature
a true fold bifurcation at a small distance from a band edge. The fold location is
independent of the size of the computational domain for these vortex families.

The paper is organized as follows. Section \ref{models} introduces the two models
which we inspect, namely the Gross--Pitaevskii equation with a periodic potential
and the discrete nonlinear Schr\"odinger equation. The connection between these
models as well as the asymptotics of gap solitons near the band edges are reviewed.
Section \ref{numerics1} gives numerical
results for the family of fundamental vortices. Section \ref{numerics2} illustrates fold
bifurcations for families of quadrupole and dipole vortex configurations. In Section \ref{conclusion}
we summarize our findings.

\section{Models}
\label{models}

The stationary Gross--Pitaevskii equation with a periodic potential in the space of two dimensions
takes the form
\beq\label{E:PNLS}
- \Delta \varphi + V(x,y) \varphi  - |\varphi|^2\varphi = \omega \varphi, \quad  (x,y) \in \RR^2,
\eeq
where the focusing case is considered, the $2\pi$-periodic potential $V$
in each coordinate is assumed to be bounded, and $\omega\in \RR$ is taken in
a spectral gap of the Schr\"{o}dinger operator
$$
L_0 := -\Delta + V.
$$
For simplicity, we assume that the periodic potential $V$ 
has even symmetries with respect to reflections
about $x = 0$ and $y = 0$.

When $\omega$ is close to the upper edge $\omega_0$ of a spectral gap of $L_0$,
a slowly varying envelope approximation of localized states can be derived and rigorously
justified in the focusing stationary Gross--Pitaevskii equation \cite{DPS09,DU09}.
In the simplest case when $\omega_0$ is attained by only one extremum of the band
structure and the Hessian at the extremum is definite, the resulting approximation is
\begin{equation}
\label{expansion}
\varphi(x,y) = \epsilon \psi(\epsilon x, \epsilon y) \varphi_0(x,y) + \mathcal{O}_{H^s}(\epsilon^{2/3}),  \quad
\omega = \omega_0 - \epsilon^2,
\end{equation}
where $s>1$ is arbitrary, $\varphi_0$ is the Bloch function at the band
edge $\omega_0$, and $\psi = \psi(X,Y)$ in slow variables $X = \epsilon x$
and $Y = \epsilon y$ satisfies the stationary nonlinear Schr\"{o}dinger (NLS) equation. 
This effective NLS equation is written in the form,
\beq\label{E:NLS}
\alpha (\psi_{XX} + \psi_{YY}) + \beta |\psi|^2\psi = \psi,
\eeq
where $\alpha > 0$ is related to the band curvature at the point $\omega_0$
and $\beta > 0$ is related to a norm of the Bloch function $\varphi_0$.
We note that the leading-order term $\epsilon \psi(\epsilon x, \epsilon y) \varphi_0(x,y)$
has the order $\mathcal{O}_{H^s}(1)$ as $\epsilon \to 0$ and hence expansion (\ref{expansion})
shows that the perturbation term is smaller than the leading-order term in the $H^s$ norm, 
where $H^s$ is Sobolev space of square integrable functions and their derivatives 
up to the $s$-th order.
When $\omega_0$ is the (upper) edge of the semiinfinite gap, the
error was shown to be $\mathcal{O}_{H^s}(\epsilon)$ or
$\mathcal{O}_{L^{\infty}}(\epsilon^2)$ \cite{Ilan}.

The main theorem of \cite{DPS09,DU09} states that if $\psi$ satisfies 
certain reversibility symmetries such as
\begin{equation}
\label{symmetry1}
\psi(X,Y) = \pm \bar{\psi}(-X,Y) = \pm \bar{\psi}(X,-Y)
\end{equation}
or
\begin{equation}
\label{symmetry2}
\psi(X,Y) = \pm \bar{\psi}(Y,X) = \pm \bar{\psi}(-Y,-X),
\end{equation}
and if the linearization of the stationary NLS equation (\ref{E:NLS}) 
is non-degenerate, then a localized solution of the Gross--Pitaevskii equation (\ref{E:PNLS})
with the asymptotic expansion (\ref{expansion}) exists in $H^s$ for this $\psi$. 
In particular, the stationary NLS equation (\ref{E:NLS}) admits 
the fundamental vortex of charge $m \in \mathbb{N}$,
\begin{equation}
\label{E:nls_vort}
\psi(X,Y) = \rho(R) e^{\ii m \theta}, \quad R = \sqrt{X^2 + Y^2}, \quad \theta = \arg(X + \ii Y),
\end{equation}
where $\rho(R) > 0$ for all $R > 0$ satisfies a certain differential equation
that follows from the stationary NLS equation (\ref{E:NLS}).
Vortex solution (\ref{E:nls_vort}) satisfies symmetry (\ref{symmetry1}) 
and the linearization of (\ref{E:NLS}) at this $\psi$ is non-degenerate. 
Hence conditions of the main theorem in \cite{DPS09,DU09} are validated and there exists 
a unique localized solutions of (\ref{E:PNLS}) continued from this $\psi$ 
with the asymptotic expansion (\ref{expansion}). Continuation of 
fundamental vortices is considered in Section \ref{numerics1}.

In the tight-binding limit of narrow spectral bands, several authors
\cite{AH09,MPS08,PS10} rigorously justified that the localized states
of the stationary Gross--Pitaevskii equation (\ref{E:PNLS})
can be described by the localized states of the stationary discrete nonlinear
Schr\"odinger (DNLS) equation,
\beq\label{E:DNLS}
-(\Delta_{\text{disc}}\phi)_{m,n} - |\phi_{m,n}|^2\phi_{m,n} = \omega \phi_{m,n},
\qquad (m,n) \in \ZZ^2,
\eeq
where
$$
(\Delta_{\text{disc}}\phi)_{m,n}=\phi_{m+1,n}+\phi_{m,n+1}+\phi_{m-1,n}+\phi_{m,n-1}-4\phi_{m,n}
$$
and $\omega \notin \sigma(-\Delta_{\text{disc}}) = [0,4]$. The DNLS equation can simplify
numerical approximations of the continuous Gross--Pitaevskii equation but does not
change properties of localized states. In particular, bifurcations of localized states are possible
from the band edge $\omega = 0$ in the focusing case. Moreover, the same
method of asymptotic multi-scale expansions can be adopted to the DNLS equation with
the expansion
\beq\label{E:SVEA}
\phi_{m,n} = \epsilon \psi(\epsilon m, \epsilon n) + o_{l^2}(1),  \quad
\omega = - \epsilon^2,
\eeq
where $\psi$ satisfies the same stationary NLS equation (\ref{E:NLS}).
A rigorous justification of the continuous NLS equation as an asymptotic 
model for ground states of the DNLS equation was recently developed 
in \cite{Bambusi1,Bambusi2}. Approximations for vortices
in this context were not obtained to the best of our knowledge.

\section{Vortex family connected to the spectral edge}
\label{numerics1}

We compute here a family of the fundamental vortices in the DNLS equation
(\ref{E:DNLS}) by using the near-edge asymptotics  \eqref{E:SVEA}
with $\psi$ given by the continuous vortex (\ref{E:nls_vort}). We will also
compare this behavior with the one in the Gross--Pitaevskii equation (\ref{E:PNLS}).

Choosing the vortex of charge one in the form (\ref{E:nls_vort}) with $m = 1$,
we compute the positive spatial profile $\rho$ by the shooting method. We let
next $\eps=\sqrt{0.03}$, so that the expansion \eqref{E:SVEA} produces an
initial guess for a solution $\phi$ of the DNLS equation (\ref{E:DNLS})
with $\omega=-0.03$ and compute $\phi$ via the Newton's method.

\begin{figure}[htpb]
  \begin{center}
    \includegraphics[scale=0.65]{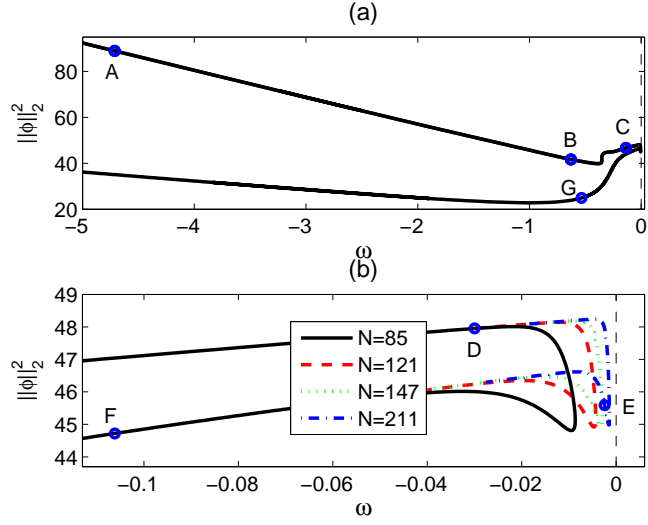}
  \end{center}
  \caption{Family of vortex solutions of \eqref{E:DNLS} continued from the vortex
  (\ref{E:nls_vort}) via the envelope approximation \eqref{E:SVEA} at $\omega=-0.03$ (point D).
  (a) A fixed computational domain is used with $N = 85$. (b) Detail of the
  vicinity of the spectral edge for a range of sizes of the computational domain.}
  \label{F:vort_fam}
\end{figure}

Next, we continue the family in the $(\omega,\|\phi\|_{l^2}^2)$-plane
using the pseudo-arclength continuation \cite{Keller77,Keller79}, in
which both $\phi$ and $\omega$ are unknowns, combined with the Newton's method. The resulting solution family
is plotted in Figure \ref{F:vort_fam}(a) using the computational domain
$[-42,42]^2\subset \ZZ^2$. The starting point at $\omega = - \eps^2 = -0.03$
is marked as $D$ in Figure \ref{F:vort_fam}(b).

The family of vortices with charge $m = 1$ seems to fold and
never reach the spectral edge contrary to the approximation \eqref{E:SVEA}.
In Figure  \ref{F:vort_fam}(b) this folding is, however, shown to be
merely a numerical artifact caused by the truncation of the infinite
domain $\ZZ^2$. The fold location approaches the edge $\omega=\omega_0=0$
as the computational domain is enlarged. The family branch containing $A-D$
thus terminates at the edge if computed on domains of diverging size.

\begin{figure}[htpb]
  %\begin{center}
    \includegraphics[scale=0.6]{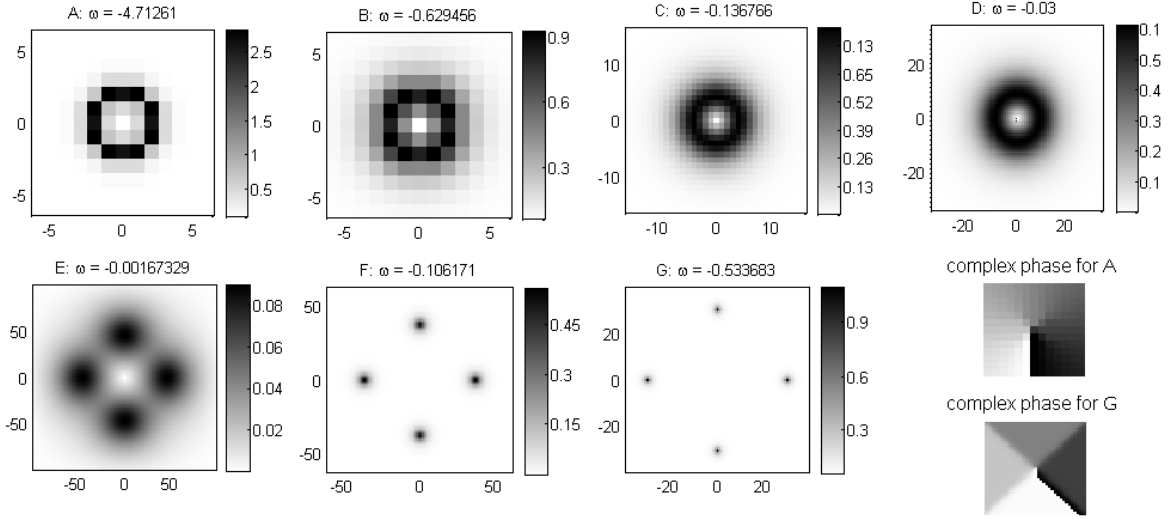}
  %\end{center}
  \caption{A-G: Modulus of the discrete vortex solutions labeled in
  Fig. \ref{F:vort_fam}. Bottom right: plots of the complex phase for vortices A and G.}
  \label{F:vort_profs}
\end{figure}

In the vicinity of the fold bifurcation, the solutions are inherent to the
truncated domain and do not correspond to any solution of
the DNLS equation \eqref{E:DNLS} on $\ZZ^2$. The fold divides the
family into two branches. Only the first branch
containing points $A,B,C,D$ is a continuation of the vortex family with
the asymptotics given by \eqref{E:SVEA} with $\psi$ as the continuous vortex \eqref{E:nls_vort}.
The other branch corresponds to a different family of vortex solutions.

Figure \ref{F:vort_profs} shows the solutions
labeled $A-G$ in Figure \ref{F:vort_fam}. The limiting behavior
for $\omega \rightarrow -\infty$ along the branch with $A$ and $B$
is a vortex with a square structure with sides of length five and
three active sites on each side, i.e. the excited sites are at
$$
(-1,-2),(0,-2),(1,-2),(-1,2),(0,2),(1,2),(-2,-1),(-2,0),(-2,1),(2,-1),(2,0),(2,1) \in \ZZ^2.
$$
Along the other branch beyond the point $G$ in the direction of decreasing $\omega$ the four solution
peaks get further localized approaching single site excitations
as $\omega \to -\infty$. The vortices keep charge one along the whole family. 
The complex phase for vortices A and G is plotted on the bottom right panel 
of Figure \ref{F:vort_profs}.

\begin{figure}[htpb]
  \begin{center}
    \includegraphics[scale=0.65]{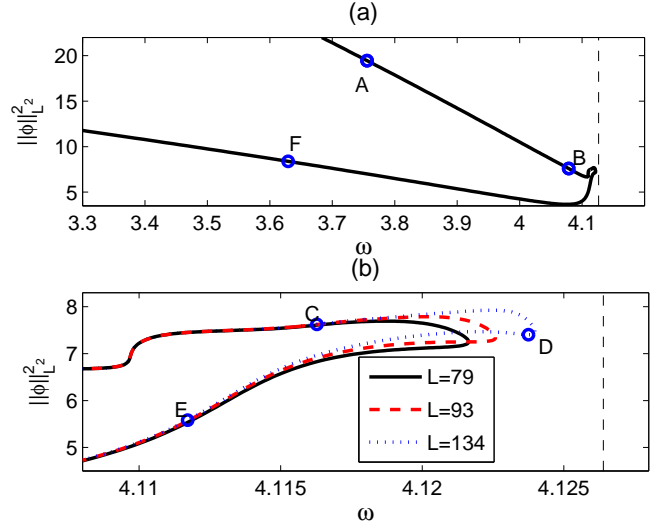}
  \end{center}
  \caption{Family of vortex solutions of \eqref{E:PNLS} continued from
  the vortex (\ref{E:nls_vort}) via the envelope approximation \eqref{E:SVEA}
  at $\omega\approx 4.09$ (point B). (a) A fixed computational domain is used with $L=93$.
  (b) Detail of the vicinity of the spectral edge for a range of sizes of the computational domain.}
  \label{F:cont_vort_fam}
\end{figure}

\begin{figure}[htpb]
  \begin{center}
    \includegraphics[scale=0.6]{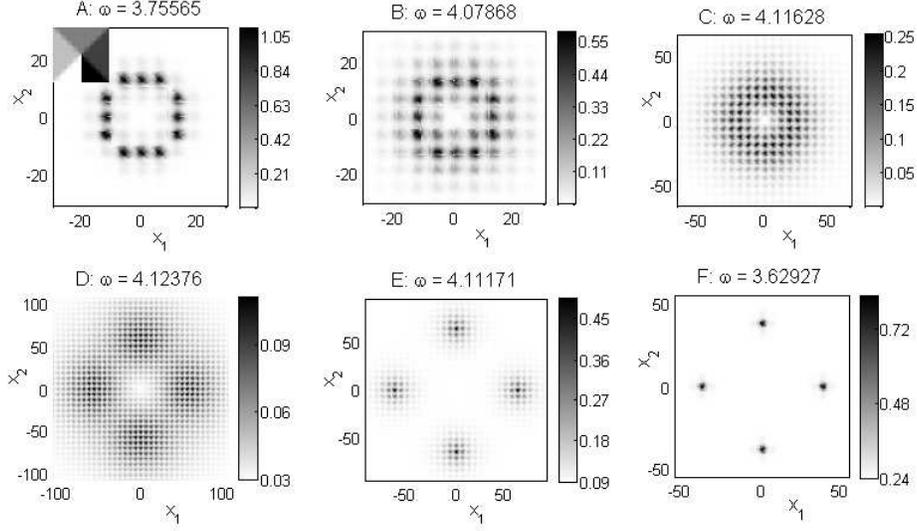}
  \end{center}
  \caption{Modulus of the continuous vortex solutions labeled in Fig. \ref{F:cont_vort_fam}. The inset in A
  shows a qualitative plot of the complex phase for all vortices A-F.}
  \label{F:contvort_profs}
\end{figure}

A similar situation arises for vortex solutions of the continuous Gross--Pitaevskii
equation \eqref{E:PNLS}. In Figures \ref{F:cont_vort_fam} and \ref{F:contvort_profs},
we present a family of vortices for
$$
V(x,y) = 6  \sin^2(x) + 6 \sin^2(y),
$$
which was the potential used in \cite{Wang}. We consider the vicinity
of the lowest spectral edge $\omega_0 \approx 4.1264$. The selected family
is qualitatively similar to the discrete vortex family above and also
terminates at the gap edge. The computational domain is $[-L/2,L/2]^2\subset \RR^2$
and the stationary equation \eqref{E:PNLS} is discretized via central difference
formulas of order 4.

The results of this section contradict the claim from \cite{Wang}
that no vortex families can be continued to the band edge of the Bloch spectrum.
These results illustrate the validity of the main theorems from \cite{DPS09,DU09},
which point out the possibility of such continuations for solutions satisfying 
reversibility symmetries (\ref{symmetry1}) or (\ref{symmetry2}) and the non-degeneracy 
conditions, e.g., for the fundamental vortex solutions.


%-------------------------------------------------------------------
\section{Quadrupole and dipole vortex families disconnected from the spectral edge}
\label{numerics2}

\begin{figure}[htpb]
  \begin{center}
    \includegraphics[scale=0.6]{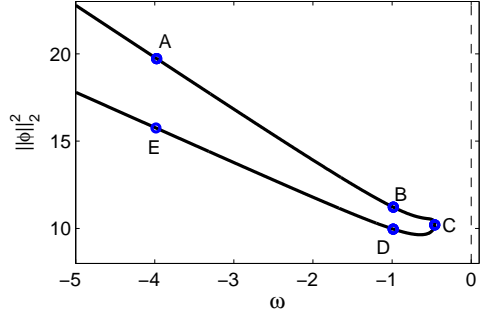}
  \end{center}
  \caption{Family of discrete vortex solutions continued from a quadrupole vortex
  and four nearest neighbor excited sites as $\omega\rightarrow -\infty$.}
  \label{F:4pole_fam}
\end{figure}

A classical example of a vortex solution of the Gross--Pitaevskii equation
\eqref{E:PNLS} and of the DNLS equation \eqref{E:DNLS} is the quadrupole vortex
with the four nearest neighbor sites excited in a square arrangement
in the asymptotics $\omega \rightarrow -\infty$, i.e. with the excited sites at
$$
(-1,0),(1,0),(0,-1),(0,1) \in \ZZ^2.
$$
For the Gross--Pitaevskii equation the excited sites are understood as
locations of the single wells (minima) of the periodic potential $V$.
We obtain this family for the DNLS equation via the initial guess
$$
\phi_{m,n} = 2 \delta_{m,1}\delta_{n,0} +
2 \ii \delta_{m,0} \delta_{n,1} - 2 \delta_{m,-1}\delta_{n,0}
- 2 \ii \delta_{m,0} \delta_{n,-1}, \quad (m,n) \in \ZZ^2,
$$
at $\omega = -1.2$ (point $D$ in Figure \ref{F:4pole_fam}), where
$\delta_{m,m_1}\delta_{n,n_1}$ is the Kronecker unit vector at
$(m_1,n_1)$ on $\ZZ^2$. The family is then continued, once again,
via the pseudo-arclength continuation combined with the Newton iteration.
\begin{figure}[htpb]
  \begin{center}
    \includegraphics[scale=0.6]{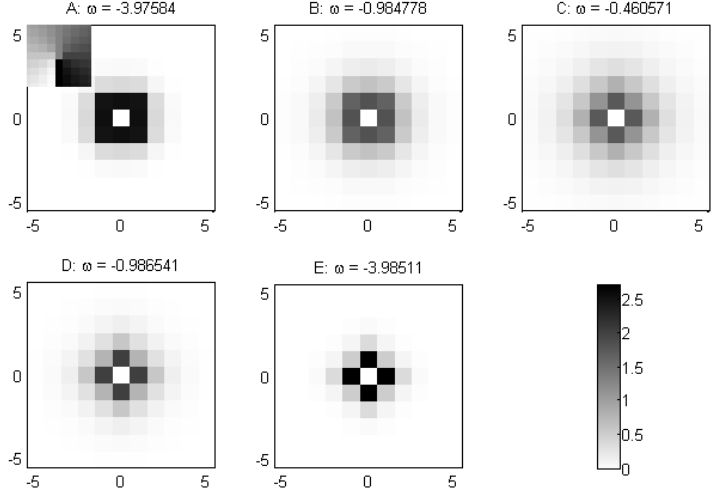}
  \end{center}
  \caption{Modulus of the discrete quadrupole vortex solutions labeled in Fig. \ref{F:4pole_fam}.
  The inset in A shows a qualitative plot of the complex phase.}
  \label{F:4pole_profs}
\end{figure}

In Figures \ref{F:4pole_fam} and \ref{F:4pole_profs}
the family curve and several solution profiles are plotted.
The computation was performed on the domains $[-N,N]^2$
with $N=10, 16, 20, 40$ and the curves in the $(\omega,\|\phi\|_{l^2}^2)$
plane remained within the distance of $\mathcal{O}(10^{-6})$
for all these $N$ and did not approach the edge.
This family folds and does not reach the spectral edge and
all its solutions remain tightly localized. It does not, therefore,
contain a slowly varying solution near the band edge, 
which could be approximated via the envelope
approximation \eqref{E:SVEA}. Qualitatively this family corresponds
to the family $a-d$ in Fig. 1(a) of \cite{Wang}.
The branch past the fold, i.e. the branch with $A$ and $B$,
has as the asymptotic profile for $\omega\rightarrow -\infty$
a square vortex with excited sites at
$$
(-1,-1),(-1,0),(-1,1),(0,1),(1,1),(1,0),(1,-1),(0,-1) \in \ZZ^2.
$$

\begin{figure}[htpb]
  \begin{center}
    \includegraphics[scale=0.6]{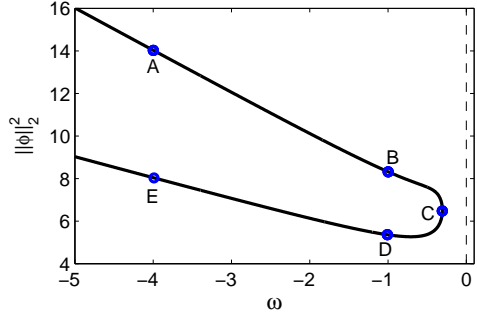}
  \end{center}
  \caption{Family of discrete vortex solutions continued from a dipole vortex
    with two nearest neighbor excited sites as $\omega\rightarrow -\infty$.}
  \label{F:2pole_fam}
\end{figure}

\begin{figure}[htpb]
  \begin{center}
    \includegraphics[scale=0.6]{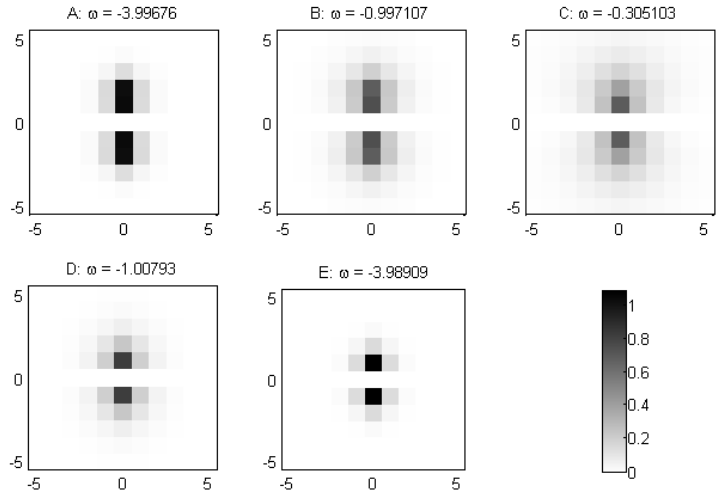}
  \end{center}
  \caption{Modulus of the discrete dipole vortex solutions labeled in Fig. \ref{F:2pole_fam}.}
  \label{F:2pole_profs}
\end{figure}

We also consider a family of real dipole solutions $\phi$ which are odd in
the $n$-index and satisfy $\phi_{m,0}=0$ for all $m\in \ZZ$. It is constructed
via the ``hand made'' initial guess
$$
\phi_{m,n} = 1.5 \delta_{m,0}\delta_{n,1} -
1.5 \delta_{m,0}\delta_{n,-1}, \quad (m,n) \in \ZZ^2,
$$
at $\omega =-1.2$ (point $D$ in Figure \ref{F:2pole_fam}). The
family is plotted in Figures \ref{F:2pole_fam} and \ref{F:2pole_profs}.
The computation was performed on the domains $[-N,N]^2 \in \ZZ^2$ with $N=10,15,20,40$
and the curves in the $(\omega,\|\phi\|_{l^2}^2)$ plane remained within the
distance of $\mathcal{O}(10^{-5})$ from each other for all these $N$ and
did not approach the edge. This family contains again only tightly
localized solitons and does not continue to the spectral edge either.
Along the branch with points $E$ and $D$ the asymptotic profile
for $\omega \rightarrow -\infty$ is the dipole with the excited
sites $(0,-1)$ and $(0,1)$. Along the other branch the asymptotic
profile is twice as broad with the excited sites at
$$
(0,-2), (0,-1), (0,1),(0,2) \in \ZZ^2.
$$
The results of this section confirm the previous numerical results from \cite{Wang}, where
all quadrupole vortex families have a fold bifurcation at a small distance
from the band edge of the Bloch spectrum. Solutions throughout our quadrupole and dipole families remain tightly localized so that they are not compatible with the slowly varying approximation (\ref{E:SVEA}).
It is, however, possible that there exist other solution families, which
have a dipole or quadrupole asymptotic form as $\omega \rightarrow -\infty$ and
which do bifurcate from the spectral edge.
The bifurcation would be guaranteed by the results of \cite{DPS09,DU09} if the continuous NLS equation \eqref{E:NLS} had dipole or quadrupole solutions, which we are not aware of.

\section{Conclusion}
\label{conclusion}

We have numerically demonstrated in both the Gross-Pitaevskii equation with a
two-dimensional periodic potential and the discrete nonlinear Schr\"odinger (DNLS)
equation that there are families of fundamental vortices bifurcating from spectral edges
of the Bloch wave spectrum. This is in agreement with the analysis in \cite{DPS09,DU09}
and in contradiction with the claim in \cite{Wang} that no vortex families continue
to a spectral edge. Our fundamental vortex families  complement the vortex families 
constructed in \cite{Wang,YangBook} which all terminate at a distance from 
the spectral edge via a fold bifurcation.

We have also investigated families of quadrupole and dipole vortex configurations
of the DNLS equation. The selected families do terminate via fold bifurcations,
which are located at a small distance from the band edge, independently of the size
of the computational domain. It is an open question whether
there exist other families of quadrupole and dipole vortex configurations, which bifurcate
from spectral edges of the Bloch wave spectrum.

{\bf Acknowledgement.} The authors thank Jianke Yang for bringing
their attention to this problem. The research of T.D. is supported by the
Research Training Group 1294 of the German Research Foundation while that
of D.P. is supported in part by the NSERC grant and the Humboldt Research Foundation.

\end{document}